\documentclass[prb,nofootinbib,twocolumn,superscriptaddress]{revtex4} 

\usepackage[dvipdfm]{graphicx}
\usepackage{dcolumn}
\usepackage{bm}
\usepackage{threeparttable} 
\usepackage{times} 
\usepackage{mathptmx}
\usepackage{lscape}
\usepackage{natbib}
\usepackage{amsmath}
\usepackage{braket}

\def\degree{\kern-.2em\r{}\kern-.3em}

\begin{document}


\title{ Crystal lattice rules disordered states }

\author{Koretaka Yuge}
\affiliation{
Department of Materials Science and Engineering,  Kyoto University, Sakyo, Kyoto 606-8501, Japan\\
}%

\begin{abstract}
Based on classical statistical thermodynamics, we develop a theoretical approach that provides new insight into how macroscopic and microscopic physical properties are bridged via crystal lattice for condensed matters. We find that in order to determine macroscopic physical properties and their temperature dependence in equilibrium disordered state, information about a few {\textit{specially selected}} microscopic states, established from geometrical characteristics of the crystal lattice, is sufficient. These special states are found to be independent of constitument elements as well as of temperature, which is in contrast to the standard conception in statistical thermodyanamics where a set of microscopic states mainly contributing to determining macroscopic physical properties depend on temperature and constituent elements. Validity and applicability of the theoretical approach is confirmed through prediction of macroscopic physical properties in practical alloys, compared with prediction by full thermodynamic simulation. The present findings provide efficient and systematic prediction of macroscopic physical properties for equilibrium disordered states based on those for special microscopic states without any information of interactions for given system.  
\end{abstract}


\maketitle
\section{Introduction}
In classical statistical thermodynamics, expected value of macroscopic property, $C$, for equilibrium system can be typically obtained by canonical average of
\begin{eqnarray}
\label{eq:c}
\overline{C} = Z^{-1}\sum_{d}C^{\left(d\right)}\exp\left(-\frac{E^{\left(d\right)}}{k_{\textrm{B}}T}\right),
\end{eqnarray}
where $Z$ is partition function, $T$ is temperature, $d$ means microscopic state that the system can take, and $E^{\left(d\right)}$ and $C^{\left(d\right)}$ are energy and physical property in state $d$, respectively. Here, we consider $\overline{C}$ as macroscopic physical property that can be obtained through canonical average, including energy, density and elastic modulus. 
When $T \to 0$, summation in Eq.~(\ref{eq:c}) can be performed for the ground states with lowest energy (e.g., ordered states below critical temperature for order-disorder transition). When $T$ increases, the system can go into disordered states (e.g., states above the critical temperature) due mainly to entropy contribution.
In the disordered states, direct estimation of $\overline{C}$ through Eq.~(\ref{eq:c}) is nontrivial since number of possible microscopic states considered astronomically increases with increase of system size. Therefore, a variety of calculation techniques have been developed to effectively address $\overline{C}$. One of the most successful techniques is Monte Carlo (MC) simulation with Metropolis algorism,\cite{metro} which samples important microscopic states mainly contributing to $\overline{C}$, and subsequent modifications have been proposed such as multihistgram method, multicanonical ensembles and entropic sampling.\cite{mc1,mc2,mc3} 
Except for the case where $Z$ can be directly estimated (e.g., transfer matrix method\cite{tm1,tm2} under special conditions), estimation of 
$\overline{C}$ generally requires information of property and energy for a large number of microscopic states ($\sim 10^{6}$ states) with larger system size ($\sim 10^{3}$ particles) even in substitutional disordered states:\cite{size} This is not a surprising fact since sufficient number of sampling points should be required to take statistical average. One of the approaches that does not require multiple states is coherent potential approximation,\cite{cpa} where it considers the average occupation of elements with the lack of information about geometrical structure. 
Another approach without use of mutiple states is high-temperature expansion,\cite{ht} which can efficiently estimate energy as well as other physical properties at high temperature. 
When we focus on substitutional disordered states such as metallic alloys or semiconductor alloys, most important condition compared with other disordered states is that possible microscopic states are mainly attributed to the underlying crystal lattice. Essential questions then naturally arise (i) how lattice bridges macroscopic and microscopic physical properties to describe equilibrium disordered states, and then, (ii) whether the lattice possesses special microscopic states out of all possible states, which determines the characteristics of equilibrium disordered states. In other words, 
we might naturally expect that macroscopic physical properties can be described by information about much fewer number of microscopic states due to existence of lattice, compared with those required by statistical thermodynamics as described above. These simple, but fundamental questions would also be considered central key to unmask universal characteristics that would be hidden in the disordered states, not only for substitutional disordered states, but also for systems where their states are described on the given lattice such as statistics in magnetic systems.\cite{is1,is2,is3,is4} 
Although the existing theoretical approaches can give accurate description for equilibrium disordered states, these have not given effective answer to the questions so far. 
We here develop a theoretical approach that can provide new insight into the above questions. 
We will reveal that, macroscopic physical properties in equilibrium disordered states can be described only by a few special microscopic states established from the lattice, and these special states are independent of constituent elements as well as of temperature: This is contrary to the standard conception in statistical physics, since it is obvious from Eq.~(\ref{eq:c}) that a set of microscopic states mainly contributing to macroscopic physical property ($\overline{C}$) strongly depends on temperature and energy (i.e., interactions given by constitunent elements). Using the present findings, we can therefore take practical advantage for efficient and systematic prediction of macroscopic physical properties for disordered states based on first-principles calculations on the lattice-derived special microscopic states.
We show concept, derivation, validity and applicability of the present findings in the followings. 

\section{Derivation and concept of the theoretical approach}
In order to address the two questions above, let us start from rewriting Eq.~(\ref{eq:c}) as 
\begin{eqnarray}
\label{eq:avec}
\overline{C} = Z^{-1}G\iint g\left(C,E\right) C\exp\left(-\frac{E}{k_{\textrm{B}}T}\right) dC dE,
\end{eqnarray}
where $g\left(C,E\right)$ represents the density of microscopic states (DOS) in which the system simultaneously takes given physical property $C$ and given energy $E$, and $G$ represents normalized constant for the integration.   
$g\left(C,E\right)$ is certainly system-dependent (i.e., depending on constituent elements as well as multibody interactions), and determining $g\left(C,E\right)$ directly leads to estimation of $\overline{C}$ that one wants to know. We first reconsider the DOS, $g$, in terms of system-independent parameters, i.e., structural parameters $\xi_{k}$. For instance, in A-B binary system, $\xi_{k}$ can be defined by
$\xi_{k} = \left< \prod_{i\in k}\sigma_{i}\right>_{\textrm{lattice}}$, where $\sigma_{i}$ is Ising-like spin specifying occupation of 
element (e.g., $\sigma_{i}=+1$ for A and $\sigma_{i}=-1$ for B) at site $i$, $\left<\quad\right>_{\textrm{lattice}}$ means average over all sites on the lattice, and $k$ is the ``figure'' whose vertices consist of lattice points (e.g., 1st nearest neighbor (1NN) pair, 2NN pair, triangle, etc.).  
With this definition, all possible microscopic states on the lattice can be uniquely specified by a set of $\xi_{k}$s (hereinafter we use $\vec{\xi}$ instead).\cite{ce} Description of the states by $\vec{\xi}$ has great advantages since energy, $E$, and other physical property, $C$, in any given microscopic state can be expressed as\cite{ce}
\begin{eqnarray}
\label{eq:ce}
E = \sum_{k}\xi_{k}\Braket{\xi_{k}|E}, \quad C = \sum_{k}\xi_{k}\Braket{\xi_{k}|C}. 
\end{eqnarray}
Here, braket denotes inner product defined as $\Braket{f|g} = \rho^{-1}\sum_{d'}f^{\left(d'\right)}g^{\left(d'\right)}$ ($\rho$ is normalized constant\cite{ce}), where summation is taken over all possible microscopic states $d'$ (i.e., all possible atomic arrangements). 
\begin{figure}
\begin{center}
\includegraphics[width=0.88\linewidth]{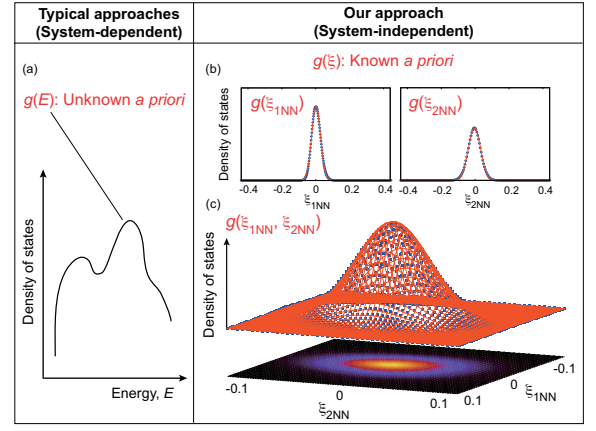}
\caption{Schematic illustration of density of microscopic states (DOS) $g$ in given system. (a) Typical approaches focus on $g$ in terms of energy, $g\left(E\right)$, which is system-dependent (unknown \textit{a priori}). (b), (c) Our approach address $g$ in terms of system-independent structural parameter 
$\xi$, $g\left(\xi\right)$ (known \textit{a priori}). $g$ for 1st nearest neighbor (1NN) and 2NN pairs (b) and their joint distribution form with the corresponding contour plot (c) are shown. 
Solid curves and closed circles denote result by analytical functions and numerical simulations for A$_{128}$B$_{128}$ binary system on fcc lattice, respectively. }
\label{fig:ndf}
\end{center}
\end{figure}
Note that Eq.~(\ref{eq:ce}) is an expansion of $E$ and $C$ with complete and orthonormal basis functions, $\xi_{k}$s, in terms of atomic arrangements.\cite{ce} 
Due to its mathematical advantages, expansion in Eq.~(\ref{eq:ce}), called cluster expansion,\cite{ce} is combined with first-principles calculations, and has been successfully applied to prediction of thermodynamic stability for alloy surface, bulk, interface and nanoparticles with multicomponent systems,\cite{ap1,ap2,ap3,ap4} and also has been extended to prediction of other properties such as elastic constants, stacking fault effects, lattice dynamics and dielectric constants for disordered as well as ordered states.\cite{ot1,ot2,ot3,ot4,TCE} 
Therefore, our following discussions based on Eq.~(\ref{eq:ce}) are general for any set of figures as well as of multibody interactions. 
Our strategy is to first give expression of DOS, $g\left(\vec{\xi}\right)$, that is system-independent but lattice-dependent, then to find landscape of $g\left(C,E\right)$ based on $g\left(\vec{\xi}\right)$, since energy and other physical properties in disordered states can be fully determined by the information of $g\left(C,E\right)$ through Eq.~(\ref{eq:avec}). 
Figure~\ref{fig:ndf} illustrates example of how our strategy starts from, based on $g\left(\xi\right)$ that can be \textit{known a priori} (Fig.~\ref{fig:ndf} (b) and (c)), while typical approaches focus on $g\left(E\right)$ that is \textit{unknown a priori} (Fig.~\ref{fig:ndf} (a)). Filled circles in Fig.~\ref{fig:ndf} (b) show the numerically estimated DOS in terms of the structural parameters for 1NN and 2NN pair, $g\left(\xi_{\textrm{1NN}}\right)$ and $g\left(\xi_{\textrm{2NN}}\right)$, and those in Fig.~\ref{fig:ndf} (c) show the DOS in terms of bivariate structural parameters for 1NN and 2NN, $g\left(\xi_{\textrm{1NN}},\xi_{\textrm{2NN}}\right)$, for fcc lattice. In order to obtain Fig.~\ref{fig:ndf} (b) and (c), we perform MC simulation to randomly sample possible states with 100000 MC step per site, using simulation box of A$_{128}$B$_{128}$ on fcc lattice. The DOSs in the figure are obtained by summing up number of states with given structural parameters over all MC steps. 
It has been shown\cite{sd} that the standard deviation of $g\left(\xi_{m}\right)$ takes $\left(ND_{m}\right)^{-1/2}$ ($N$: number of atoms in the system. $D_{m}$: number of figure $m$ per site) at equiatomic composition, and the solid curves in Fig.~\ref{fig:ndf} (b) are the analytical normal distribution functions (NDF) satisfying the corresponding standard deviation. We can see successful agreement of the DOS by the NDF for both 1NN and 2NN pairs. This tells that landscape of the DOS certainly reflects geometrical characteristics of underlying lattice. 
Solid lines in Fig.~\ref{fig:ndf} (c) is the bivariate NDF (BNDF), which also enables successful fitting to the dotted points. We confirmed that DOS, $g\left(\xi_{k}\right)$, can be well described by NDF for any type of figure $k$, and $g\left(\xi_{k},\xi_{k'}\right)$ by BNDF for any given figures of $k$ and $k'$. 
We emphasize here that while NDF commonly appears in elementary statistical physics such as spatial distribution of free particles in a rigid box derived from probability theory, the present case of $g\left(\xi_{k}\right)$ and $g\left(\xi_{k},\xi_{k'}\right)$ having NDF or BNDF does not simply relates to such case, since constituent figures on the lattice can share their vertex, edge or face with each other. 
Another important characteristic is that for any given $k$ and $k'$, the correlation coefficient, $R_{kk'}$ satisfies $\left|R_{kk'}\right| \ll 1$ (for $k=k'$, $R$ should always be 1), which means that correlation coefficient matrix, $\mathbf{R}$, for $\left\{\xi_{k}\right\}$ becomes numerically diagonal with large $N$ ($N$: number of atoms in the system): There is almost no correlation between any given two figures, although structural parameters themselves are not statistically independent.\cite{ops} These remarked properties are not confined to the fcc lattice, and we confirmed that they also hold for other representative lattices, including bcc, hcp, and diamond. Therefore, when we describe DOS, $g$, in terms of a set of $\xi_{k}$ ($k=1\ldots q$) by $q$-dimensional multivariate NDF (MNDF), $g\left(\xi_{1},\ldots , \xi_{q}\right)$, it can be decomposed into the product of individual NDF for figure $k$, $g\left(\xi_{k}\right)$: 
\begin{eqnarray}
\label{eq:mndf}
g\left(\xi_{1},\ldots ,\xi_{q}\right) \simeq \prod_{k}g\left(\xi_{k}\right),
\end{eqnarray}
since $\mathbf{R}$ is again approximately diagonal. 
Our derivation below is based on Eq.~(\ref{eq:mndf}).

In order to address DOS, $g\left(C,E\right)$, we introduce new 
variables, $\left\{\zeta_{k}\right\}$, defined by
\begin{eqnarray}
\label{eq:zeta}
\zeta_{k} = \xi_{k} / \left<\xi_{k}\right>_{\textrm{sd}}.
\end{eqnarray}
Here, $\left<\quad\right>_{\textrm{sd}}$ means standard deviation (SD), and thus, SD of $\left\{\zeta_{k}\right\}$ is normalized to 1, where 
$\left\{\zeta_{k}\right\}$ can be also described by MNDF.
To obtain MNDF with a different set of $q$ variables, $\left\{\eta_{k}\right\}$, we can introduce matrix $\mathbf{A}$ with $\vec{\eta} = \mathbf{A}\vec{\zeta}$, leading to
\begin{eqnarray}
\label{eq:feta}
g\left(\vec{\eta}\right) = \Psi\exp\left\{-\frac{1}{2}\left(\vec{\eta}-\mathbf{A}\vec{\left<\zeta\right>}\right)'\left(\mathbf{A}\mathbf{A'}\right)^{-1}\left(\vec{\eta}-\mathbf{A}\vec{\left<\zeta\right>}\right)\right\} 
\end{eqnarray}
where $\Psi = \left(\sqrt{2\pi}\right)^{-q}/\left|\mathbf{A}\right|$, $\left<\quad\right>$ means the average over all possible microscopic states, and covariance matrix for $\left\{\eta_{k}\right\}$, $\Gamma$, corresponds to $\mathbf{A}\mathbf{A'}$. When we consider matrix $\mathbf{A}$ taking 
\begin{eqnarray}
\label{eq:A}
\mathbf{A} = \left(
\begin{array}{ccc}
\Braket{\xi_{1}|E}\left<\xi_{1}\right>_{\textrm{sd}} &  \cdots & \Braket{\xi_{q}|E}\left<\xi_{q}\right>_{\textrm{sd}} \\
\Braket{\xi_{1}|C}\left<\xi_{1}\right>_{\textrm{sd}} &  \cdots & \Braket{\xi_{q}|C}\left<\xi_{q}\right>_{\textrm{sd}} \\
 & \multicolumn{1}{c}{\mathbf{B}} & \\
\end{array}
\right)
\end{eqnarray}
for any given submatrix $\mathbf{B}$ satisfying the existence of $\mathbf{A}^{-1}$, 
subspace matrix of $\Gamma$ for $\eta_{1}$ and $\eta_{2}$, $\Gamma'$, always takes
\begin{eqnarray}
\label{eq:cov}
\Gamma' = \left(
\begin{array}{lll}
\displaystyle{\sum_{k}}\Braket{\xi_{k}|E}^{2}\left<\xi_{k}\right>_{\textrm{sd}}^{2} & \displaystyle{\sum_{k}}\Braket{\xi_{k}|E}\Braket{\xi_{k}|C}\left<\xi_{k}\right>_{\textrm{sd}}^{2} \\
\displaystyle{\sum_{k}}\Braket{\xi_{k}|C}\Braket{\xi_{k}|E}\left<\xi_{k}\right>_{\textrm{sd}}^{2} & \displaystyle{\sum_{k}}\Braket{\xi_{k}|C}^{2}\left<\xi_{k}\right>_{\textrm{sd}}^{2} \\
\end{array}
\right) 
\end{eqnarray}
for a set of all possible $k$.
From above equations, it is now obvious that $\eta_{1}=E$ and $\eta_{2}=C$. It can be easily shown that $\Gamma'$ is a covariance matrix for $E$, and $C$ when $\mathbf{R}$ is diagonal. From the fact that any marginal distribution in MNDF is also the MNDF, $g\left(C,E\right)$ can be well described by BNDF. 
We therefore reveal that DOS in terms of energy and other physical property, $g\left(C,E\right)$, universally takes BNDF with covariance matrix of $\Gamma'$ in Eq.~(\ref{eq:cov}) for any constituent elements (i.e., any interactions in the system) as well as for any lattices. 
This directly means that macroscopic and microscopic physical properties for equilibrium disordered states are bridged through DOS of $g\left(C,E\right)$ given by BNDF, where its landscape (i.e., covariance matrix of $\Gamma'$) certainly reflects geometrical characteristic of the cystal lattice since $\left<\xi_{m}\right>_{\textrm{sd}}$ is determined from number of figure $m$ contained in the lattice as described above.

Since DOS, $g\left(C,E\right)$, takes the form of BNDF, the question, existence of special microscopic states to determine macroscopic physical properties in equilibrium disordered states, become identical to the question of whether special states exist to determine concrete landscape of the DOS. Our strategy is therefore to find conditions that structural parameters for special states should satisfy, to determine parameters of DOS: $\left<E\right>$, $\left<C\right>$, $\left<E\right>_{\textrm{sd}}$, $\left<C\right>_{\textrm{sd}}$ and the correlation coefficient between $C$ and $E$, $R_{CE}$.  
We first focus on average, $\left<E\right>$: From Eq.~(\ref{eq:ce}), we can immediately describe $\left<E\right> = \sum_{k}\left<\xi_{k}\right>\Braket{\xi_{k}|E}$, since coefficients, $\Braket{\xi_{k}|E}$s, are essentially independent of the microscopic states.\cite{ce} Therefore, when a single microscopic state has structural parameters of $\left\{\left<\xi_{1}\right>, \cdots, \left<\xi_{q}\right>\right\}$, the corresponding energy and other physical property respectively takes  $\left<E\right>$ and $\left<C\right>$. This state is known as special quasirandom structure,\cite{sqs,sa1} which is one of the special states. 
Existence of other special states to represent standard deviations or correlation coefficient is unknown, which should be revealed in the present study.
For simplicity (without lack of generality), hereinafter we define structural parameter $\xi_{k}$, energy $E$ and other physical property $C$ measured from their average values.
In order to determine $\left<E\right>_{\textrm{sd}}$, $\left<C\right>_{\textrm{sd}}$ and $R_{CE}$, we introduce following matrix $\Lambda$:
\begin{eqnarray}
\label{eq:lambda}
\Lambda &=& \left(
\begin{array}{cc}
 \lambda_{CC} & \lambda_{CE} \\
 \lambda_{EC} & \lambda_{EE} \\
\end{array}
\right) \nonumber \\
\lambda_{IJ} &=& \sum_{j=1}^{n_{j}}\left[\prod_{u=I,J}\left\{\sum_{t}S_{jt}\chi_{ut}^{\left(j\right)}\right\}\right], 
\end{eqnarray}
where $I$ and $J$ corresponds to property $C$ or energy $E$. $\chi_{u\beta}^{\left(l\right)} = \Braket{\xi_{\beta}|E}\xi_{\beta}^{\left(l\right)}$ when $u$ denotes $E$ and $\chi_{u\beta}^{\left(l\right)} = \Braket{\xi_{\beta}|C}\xi_{\beta}^{\left(l\right)}$ when $u$ denotes $C$ with $l=1\cdots n_{j}$ ($l$ denote structure), and $\beta$ denote figure whose number of vertex is not less than two. 
$S_{jt}$ is the element of matrix $\mathbf{S}$, which can be determined when we take $\mathbf{S}$ as the submatrix of matrix $\mathbf{H}$ with first $n_{j}$ lines, obtained through the following expression:
\begin{eqnarray}
\mathbf{H} = \lim_{p\to\infty}\hat{K}^{p}\mathbf{F}\quad \left(p\ge1\right), 
\end{eqnarray}
where $\mathbf{F}$ is the $2\times 2$ matrix in which one of the four elements takes -1 and the other three all take +1, and operator $\hat{K}$ satisfies
\begin{eqnarray}
\hat{K}^{r}\mathbf{F} &=& \left(
\begin{array}{cc}
 F_{11}\cdot\hat{K}^{\left(r-1\right)}\mathbf{F} & F_{12}\cdot\hat{K}^{\left(r-1\right)}\mathbf{F} \\
 F_{21}\cdot\hat{K}^{\left(r-1\right)}\mathbf{F} & F_{22}\cdot\hat{K}^{\left(r-1\right)}\mathbf{F} \\
\end{array}
\right) \quad \left(r\ge 1\right) \nonumber \\
\hat{K}^{0}\mathbf{F} &=& \mathbf{F}.
\end{eqnarray}
Let us consider the case in which the structure parameter of figure $k$ for structure $l$, $\xi_{k}^{\left(l\right)}$, satisfies
\begin{eqnarray}
\label{eq:alpha}
\gamma_{l}\xi_{k}^{\left(l\right)} = S_{lk} \left<\xi_{k}\right>_{\textrm{sd}}
\end{eqnarray}
for all given $k$ and $l$ considered ($\gamma_{l}$ is arbitraly positive real number). The definition of Eqs.~(\ref{eq:lambda}) and (\ref{eq:alpha}) immediately gives
\begin{eqnarray}
\label{eq:var}
\left<E\right>_{\textrm{sd}} &\simeq& \sqrt{\Gamma'_{11}} = \sqrt{\Lambda_{22}/n_{j}} = \sqrt{\sum_{l}\left\{\gamma_{l}E^{\left(l\right)}\right\}^{2}/n_{j}} \nonumber\\
\left<C\right>_{\textrm{sd}} &\simeq& \sqrt{\Gamma'_{22}} = \sqrt{\Lambda_{11}/n_{j}} = \sqrt{\sum_{l}\left\{\gamma_{l}C^{\left(l\right)}\right\}^{2}/n_{j}} \nonumber\\
R_{CE} &\simeq& \Lambda_{12} / \left(n_{j}\sqrt{\Gamma'_{11}\Gamma'_{22}}\right) \nonumber \\
&=& \sum_{l}\left\{\gamma_{l}^{2}E^{\left(l\right)}C^{\left(l\right)}/n_{j}\right\}/\left(\left<E\right>_{\textrm{sd}}\left<C\right>_{\textrm{sd}}\right) \nonumber \\
\quad
\end{eqnarray}
under the condition that $\mathbf{R}$ is approximately diagonal, since all cross terms for $\left\{\chi_{ut}\right\}$ considered can be factored out regardless of the signs of $\left\{\Braket{\xi_{\beta}|E}\right\}$ and $\left\{\Braket{\xi_{\beta}|C}\right\}$. 
This indicates that when a set of special states $\left\{l\right\}$ with structural parameters $\left\{\xi_{k}^{\left(l\right)}\right\}$ satisfying Eqs.~(\ref{eq:lambda}) and (\ref{eq:alpha}) are found, standard deviation of energy and other physical property, and correlation coefficients of $R_{CE}$, can be determined only by information (i.e., $E^{\left(l\right)}$ and $C^{\left(l\right)}$) about lattice-derived special microscopic states through Eq.~(\ref{eq:var}). From Eqs.~(\ref{eq:lambda})-(\ref{eq:var}), it is now obvious that these special microscopic states are independent of constituent elements as well as of temperature, since they can be constructed by infomation only about lattice-dependent parameters, $\left<\xi_{k}\right>_{\textrm{sd}}$s.
\begin{figure}
\begin{center}
\includegraphics[width=0.71\linewidth]{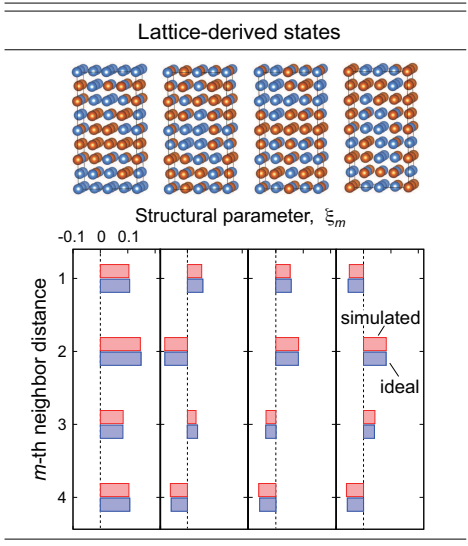}
\caption{Simulated atomic arrangements of lattice-derived special states for equiatomic composition in binary system on fcc lattice. 
Red and blue bars denote simulated and ideal values of structural parameters for up to 4th neighbor distance, respectively.  }
\label{fig:sp}
\end{center}
\end{figure}

From the above discussions, we reveal that lattice-derived special states can exist to determine macroscopic energy and other physical properties, and their temperature dependence for equilibrium disordered states. Based on Eq.~(\ref{eq:avec}), we can directly describe macroscopic physical properties using those for the special microscopic states, ``without'' any information of interactions in the system, since DOS, $g\left(C,E\right)$, can be again determined by information about the special states as discussed above. 
We here point out that even if one finds conditions where correlation coefficients of DOSs for given two figures do not approach to zero with the increase of system size (i.e., correlation coefficient matrix, $\mathbf{R}$, is not diagonal with the structural parameters, $\xi$s), our theoretical approach still holds for such systems. In this case, we would employ another set of structural parameters to give diagonal matrix of $\mathbf{R}$. One straightforward approach is to find the eigenvectors for $\mathbf{R}$, $\overline{\xi}$s. Since $\overline{\xi}$s are linear combination of $\xi$s (e.g., $\overline{\xi}_{k'}=\sum_{k}a_{k'k}\xi_{k}$), DOS in terms of the eigenvectors, $g\left(\overline{\xi}_{1},\ldots,\overline{\xi}_{q}\right)$, can also be described by $q$-dimensional NDF, which satisfies $g\left(\overline{\xi}_{1},\ldots,\overline{\xi}_{q}\right) \simeq \prod_{k'=1}^{q}g\left(\overline{\xi}_{k'}\right)$. With these eigenvectors, physical property and energy can merely be re-described by $C = \sum_{k'}\overline{\xi}_{k'}\Braket{\overline{\xi}_{k'}|C}$ and $E = \sum_{k'}\overline{\xi}_{k'}\Braket{\overline{\xi}_{k'}|E}$. Using these new basis functions of $\xi_{k'}$s, we can describe DOS by MNDF with diagonal correlation coefficient matrix, which is required by the present theoretical approach. 
With these considerations, the present approach does not essentially require the condition of diagonal $\mathbf{R}$ depending on coordination of configuration space, but merely requires that DOS for majority of microscopic states can be described by MNDF. In fact, we confirm that this requirement for DOS widely holds for binary as well as multicomponent system in crystalline materials, by using computer simulation of the DOS in terms of structural parameters for multicomponent system on representative lattices (e.g., including fcc, bcc, hcp and diamond lattice). 
We also note here that the present theoretical approach for describing disordered states is essentially different from high-temperature expansion. Let us consider the simple case when the system is described by Ising model with nearest-neighbor interaction, $\mu$. Applicability of the high-temperature expansion relies on magnitude relationship between $\mu$ and $k_{{B}}T$, i.e., $k_{{B}}T \gg \left|\mu\right|$. Meanwhile, in the present approach, we make no assumption of $k_{{B}}T \gg \left|\mu\right|$. We make single approximation that DOS in terms of structural parameters is described by Eq.~(\ref{eq:mndf}). Since the DOS is independent of $T$ or $\mu$, applicability of the present approach relies on lattice, i.e., how well the DOS is described by Eq.~(\ref{eq:mndf}) for given lattice. Therefore, when the DOS near as well as far from its center of gravity is well described by Eq.~(\ref{eq:mndf}) for given lattice, the present approach gives successful description of energy and property not only at high temperatures but also at low temperatures. With this consideration, the present theoretical approach is completely different from the high-temperature expansion.

\section{Application to the practical systems}
In order to practically obtain the special states for given lattice, we show example to search minimal number of special states in A-B binary system at equiatomic composition on fcc lattice. 
The special states can contain information of any type of figure (i.e., short as well as long range correlations) when a set of structural parameters satisfies the above conditions. We here perform MC simulation to minimize difference of structural parameters between simulated values of $\left\{\xi_{k}^{\left(\textrm{MC}\right)}\right\}$ and ideal values of $\left\{\xi_{k}^{\left(0\right)}\right\}$, defined by $\sum_{k}\left(\xi_{k}^{\left(\textrm{MC}\right)}-\xi_{k}^{\left(0\right)}\right)^{2}$, where figure $k$ is confined to 1NN-4NN pairs and triangles having at least one 1NN pair and number of atoms in the MC simulation is up to 64. The results are summarized in Fig.~\ref{fig:sp}, where red and blue bars represent simulated and ideal structural parameters of 1NN-4NN pairs. 
We here emphasize again that to construct special states, we do not require any information about constituent elements or about temperature. 
We can see successful agreement of simulated structural parameters with ideal ones, indicating that the special states can be numerically constructed within a limited system size.  
\begin{figure}
\begin{center}
\includegraphics[width=1.00\linewidth]{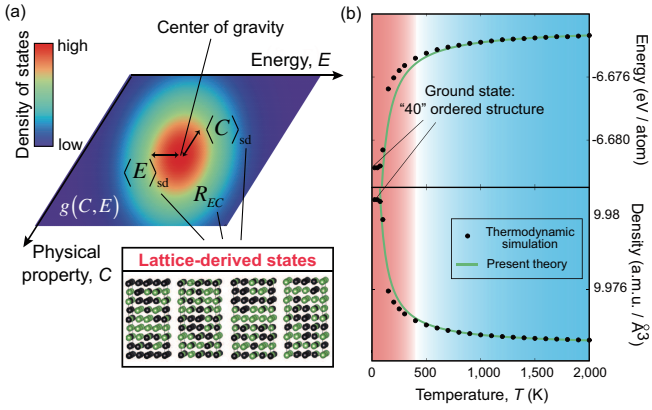}
\caption{(a) Schematic illustration of landscape of DOS in terms of energy and other physical properties, $g\left(C,E\right)$ (contour plot). Temperature dependence of their values in disordered states are fully determined by the knowledge of $g\left(C,E\right)$. Landscape of $g\left(C,E\right)$ can be determined by the information of the lattice-derived special states. (b) Temperature dependence of macroscopic internal energy and density for Pt$_{50}$Rh$_{50}$ binary system. Closed circles denote results of full thermodynamic approach using MC simulation and cluster expansion. Solid curves denote results by the present theoretical approach using information of the lattice-derived special states.  }
\label{fig:tc}
\end{center}
\end{figure}

Validity and applicability of the present theoretical approach should then be demonstrated through prediction of energy and physical property (here, density) in disordered states and their temperature dependence using information of special states shown in Fig.~\ref{fig:sp}. 
In the present study, Pt$_{50}$Rh$_{50}$ binary alloy is chosen as model system. 
In Fig.~\ref{fig:tc} (a), we show schematic illustration of universal landscape of DOS in terms of energy and other physical property, by using contour plot of $g\left(C,E\right)$, which is in the form of BNDF. We also show how the landscapes of the DOS is ruled by the lattice-derived special states. 
We first predict macroscopic energy and density of Pt$_{50}$Rh$_{50}$ binary alloy in equilibrium state, based on the present theoretical approach: (i) energy and density for four special microscopic states shown in Fig.~\ref{fig:sp} and single special quasirandom structure\cite{sqs} for Pt$_{50}$Rh$_{50}$ alloy are estimated by first-principles calculation (details of first-principles calculation is described later), (ii) these energy and density are applied to Eqs.~(\ref{eq:var}) to determine $\left<E\right>_{\textrm{sd}}$, $\left<C\right>_{\textrm{sd}}$ and $R_{CE}$, which directly determine DOS, $g\left(C,E\right)$, and (iii) obtained DOS, $g\left(C,E\right)$ is applied to Eq.~(\ref{eq:avec}) to estimate macroscopic energy and density.
The results predicted by the present theoretical approach are compared with full thermodynamic approach in a state-of-the-art manner, based on MC simulation and cluster expansion (CE). In the CE, coefficients, $\Braket{\xi_{k}|E}$s and $\Braket{\xi_{k}|C}$s, are determined by fitting the DFT results of 301 structures to Eq.~(\ref{eq:ce}), which consists of up to 32 atoms with a variety of atomic configurations and compositions. These structures are iteratively obtained by CE and first-principles calculations in order to accurately describe structures with high and low energies.\cite{ap3,st1} 
In the first-principles calculation, total energy and density are estimated by the VASP code\cite{vasp1, vasp2} based on the projector-augmented wave method\cite{paw} within the generalized-gradient approximation of Perdew-Burke-Ernzerhof (GGA-PBE)\cite{pbe} to the exchange-correlation functional. 
The plane wave cutoff of 360 eV is used, and atomic positions are relaxed until the residual forces become less than 0.001 eV/angstrom. 
We obtained twelve (consisting of one empty, one on-site, six pair up to sixth nearest neighbor, three triplet and one quartet figure) optimized coefficients with prediction accuracy, a cross-validation score,\cite{cv} of 0.5 meV/atom, which gives sufficient accuracy to capture the thermodynamics for Pt$_{50}$Rh$_{50}$ alloys. These coefficients are applied to MC simulation under a canonical ensemble with a simulation box having 4000 atoms (2000 Pt and 2000 Rh atoms on fcc lattice) to take the statistical average (8000 MC step per site) of energy  and density at each temperature, where the simulated results are shown in Fig.~\ref{fig:tc} (b). 
Full thermodynamic simulation (closed circles) exhibits order-disorder transition below 200 K, which can be seen by discontinuous changes in energy. 
The predicted ground-state at $T=0$ K for Pt$_{50}$Rh$_{50}$ is ``40'', which agrees with previous theoretical works.\cite{fp1, fp2}
Near and below the order-disorder transition temperature (red-colored area in Fig.~\ref{fig:tc} (b)), deviation between the theory and MC results can be seen. 
Note that this deviation does not come from the fact that special states in Fig.~\ref{fig:sp} do not include contribution of figures other than the considered figures: e.g., from the present theory, the special states give standard deviation of energy of 0.563 eV, which exhibits successful agreement with thermodynamic  simulation result of 0.559 eV for the Pt$_{50}$Rh$_{50}$ alloys. We confirm that if we further include longer-range or higher-dimensional figures into special states to modify the predicted standard deviation or covariance in Eq.~(\ref{eq:var}), results in Fig.~\ref{fig:tc} (b) show no significant change.
This deviation should reflect the fact that DOS far from the center of gravity in terms of structural parameters would not take the form of MNDF for fcc lattice due to the statistical interdependence of constituent figures on lattice as described above.  
However, when temperature increases from the transition temperature (blue-colored area in Fig.~\ref{fig:tc} (b)), the present theory exhibits excellent agreement with all MC results of  energy and other property.  
This should be an important fact since: Full thermodynamic simulation indicates that ``with'' explicit information of interactions, $\sim30$ million states is required to predict macroscopic physical properties for equilibruim disordered states, while the present study reveals that information only about lattice-derived special states (as shown in Fig.~\ref{fig:sp}) are sufficient ``without'' any information of interactions for the system. This certainly demonstrates that macroscopic physical properties in equilibrium disordered states can be reasonablly determined from a few special microscopic states established from the lattice.  

\section{Conclusions}
We develop a theoretical approach that provides new insight into how crystal lattice bridges macroscopic and microscopic physical properties for equilibrium disordered states, through findings of representation of density of microscopic states in terms of energy and other physical properties. We find that a few special microscopic states established from lattice is sufficient to determine macroscopic physical properties and their temperature dependence. These special states are independent of temperature as well as constituent elements, which is a natural outcome that landscape of the density of states significantly reflects geometrical characteristics of the lattice. The present findings enable efficient prediction of macroscopic physical properties based on first-principles calculation, where its validity and applicability is confirmed for Pt-Rh binary alloys through prediction of their energy and physical property (density), compared with the result by the full thermodynamic approach based on cluster expansion and Monte Carlo simulation. 

\section*{Acknowledgements}
This work was supported by a Grant-in-Aid for Scientific Research on Innovative Areas ``Materials Science on Synchronized LPSO Structure'' (26109710) and a Grant-in-Aid for Young Scientists B (25820323) from the MEXT of Japan, Research Grant from Hitachi Metals$\cdot$Materials Science Foundation, and Advanced Low Carbon Technology Research and Development Program of the Japan Science and Technology Agency (JST). 


\begin{thebibliography}{9}
\bibitem{metro} N. Metropolis, A.W. Rosenbluth, M.N. Rosenbluth, A.H. Teller, and E. Teller, J. Chem. Phys. {\textbf{21}}, 1087 (1953).
\bibitem{mc1} A.M. Ferrenberg and R. H. Swendsen, Phys. Rev. Lett. \textbf{63}, 1195 (1989).
\bibitem{mc2} G. Bhanot, R. Salvador, S. Black, P. Carter, and R. Toral, Phys. Rev. Lett. \textbf{59}, 803 (1987).
\bibitem{mc3} J. Lee, Phys. Rev. Lett. \textbf{71}, 211 (1993).
\bibitem{tm1} P.A. Pearce, Phys. Rev. Lett. \textbf{58}, 1502 (1987).
\bibitem{tm2} W. Guo, B. Nienhuis and H.W.J. Bl\"{o}te, Phys. Rev. Lett. \textbf{96}, 045704 (2006).
\bibitem{size} K. Binder, Phys. Rev. Lett. \textbf{45}, 811 (1980).
\bibitem{cpa} M. Jaros, Rep. Prog. Phys. \textbf{48}, 1091 (1985).
\bibitem{ht} F.Y. Wu, Rev. Mod. Phys. \textbf{54}, 235 (1982). 
\bibitem{is1} D. Winter, P. Virnau, and K. Binder, Phys. Rev. Lett. \textbf{103}, 225703 (2009).
\bibitem{is2} T. Nakamura, Phys. Rev. Lett. \textbf{101}, 210602 (2008).
\bibitem{is3} Y.L. Loh and E.W. Carlson, Phys. Rev. Lett. \textbf{97}, 227205 (2006).
\bibitem{is4} F. Zhou, T. Maxisch, and G. Ceder, Phys. Rev. Lett. \textbf{97}, 155704 (2006).
\bibitem{ce}J.M. Sanchez, F. Ducastelle, and D. Gratias, Physica {\textbf{128A}}, 334 (1984).
\bibitem{ap1} A.V. Ruban and H.L. Skriver, Comput. Mater. Sci. {\textbf{15}}, 119 (1999). 
\bibitem{ap2} S. M\"{u}ller, M. St\"{o}hr, and O. Wieckhorst, Appl. Phys. A {\textbf{415}}, 82 (2006). 
\bibitem{ap3} K. Yuge, Phys. Rev. B {\textbf{84}}, 085451 (2011).
\bibitem{ap4} F. Lechermann, M. F\"{a}hnle, and J.M. Sanchez, Intermetallics {\textbf{13}}, 1096 (2005).
\bibitem{ap5} K. Yuge, Phys. Rev. B {\textbf{85}}, 144105 (2012).
\bibitem{ot1} G.D. Garbulsky and G. Ceder, Phys. Rev. B {\textbf{53}}, 8993 (1996). 
\bibitem{ot2} A. van de Walle and G. Ceder, Rev. Mod. Phys. {\textbf{74}}, 11 (2002).
\bibitem{ot3} K. Yuge, J. Phys.: Condens. Matter {\textbf{21}}, 415403 (2009).
\bibitem{ot4} K. Yuge, R. Saito and J. Kawai, Phys. Rev. B {\textbf{87}}, 024105 (2013).
\bibitem{TCE}  A. van de Walle, Nature Mater. {\textbf{7}}, 455 (2008). 
\bibitem{sd} S.-H. Wei, L.G. Ferreira, J.E. Bernard, and A. Zunger, Phys. Rev. B {\textbf{42}}, 9622 (1990).
\bibitem{ops} F. Ducastelle, {\textit{Order and Phase Stability in Alloys}} (Elsevier Science, New York, 1994).
\bibitem{sqs} A. Zunger, S.-H. Wei, L.G. Ferreira, and J.E. Bernard, Phys. Rev. Lett. {\textbf{65}}, 353 (1990).
\bibitem{sa1} K.C. Hass, L.C. Davis, and A. Zunger, Phys. Rev. B {\textbf{42}}, 3757 (1990). 
\bibitem{st1} K. Yuge, Phys. Rev. B {\textbf{84}}, 134207 (2011).
\bibitem{vasp1}G. Kresse and J. Hafner, Phys. Rev. B {\textbf{47}}, R558 (1993).
\bibitem{vasp2}G. Kresse and J. Furthm\"uller, Phys. Rev. B {\textbf{54}}, 11169 (1996).
\bibitem{paw}G. Kresse and D. Joubert, Phys. Rev. B {\textbf{59}}, 1758 (1999). 
\bibitem{pbe} J.P. Perdew, K. Burke, and M. Ernzerhof, Phys. Rev. Lett. {\textbf{77}}, 3865 (1996).
\bibitem{cv} A. van de Walle, M. Asta, and G. Ceder, Calphad {\textbf{26}}, 539 (2002).
\bibitem{fp1} Z.W. Lu, B.M. Klein, and A. Zunger, J. Phase Equilib. {\textbf{16}}, 36 (1995).
\bibitem{fp2} J. Pohl and K. Albe, Acta Mater. {\textbf{57}}, 4140 (2009).
\end{thebibliography}
\end{document}